\documentclass[10pt,letterpaper]{article}
\usepackage{opex3}
\usepackage{amsmath,amsfonts,amssymb,bm}\interdisplaylinepenalty=2500

\begin{document}

\title{Coherent resonant K$_{\rm a}$--band photonic microwave receiver}

\author{Vladimir S. Ilchenko, Jerry Byrd, Anatoliy A. Savchenkov, \\ David Seidel, Andrey B. Matsko,
and Lute Maleki}
\address{OEwaves Inc., 1010 East Union St., Pasadena, CA 91106}
\email{Andrey.Matsko@oewaves.com}
\date{\today}

\begin{abstract}
We propose theoretically and demonstrate experimentally a coherent
microwave photonic receiver operating at 35~GHz carrier frequency.
The device is based on a lithium niobate or lithium tantalate
optical whispering gallery mode resonator coupled to a microwave
strip line resonator. Microwave local oscillator is fed into the
microwave resonator along with the microwave signal. We show that
the sensitivity of this receiver significantly exceeds the
sensitivity of the incoherent quadratic receiver based on the same
technology. The coherent receiver can possess a dynamic range in
excess of 100 dB in 5~MHz band if a low noise laser is utilized.
\end{abstract}




\section{Introduction}

Direct microwave receivers based on  all-resonant interaction of
light and microwaves in  solid-state whispering gallery mode (WGM)
resonators have been recently developed
\cite{cohen01el-a}-\cite{ilchenko08ptl}. The sensitivity of such
devices does not degrade with the increasing microwave frequency.
Phase insensitive (direct) $X-$, $K_{\rm u}-$, and K$_{\rm a}-$band
WGM-based receivers have been demonstrated previously
\cite{cohen01el-a}-\cite{ilchenko08ptl}. In this paper we present
results of our experimental and theoretical study of phase
sensitive (coherent) WGM-based microwave photonic receivers
operating in the $K_{\rm a}-$ band. We show that such a device has
a lot of promise compared with its phase insensitive analogies. We
first briefly describe the results of our experimental studies, and
then address the theoretical issues related to the coherent
receiver.

\begin{figure}[ht]
   \begin{center}
   \begin{tabular}{c}
   \includegraphics[height=4.5cm,angle=0]{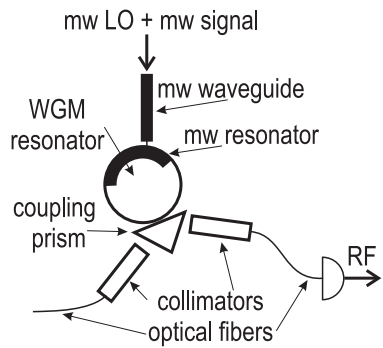}
   \end{tabular}
   \end{center}
   \caption[figs1]
{ \label{fig1}  Scheme of the coherent receiver. }
   \end{figure}

The receiver operates due to three-wave mixing occurring in the
resonator host medium possessing a quadratic nonlinearity. A
microwave signal along with a microwave local oscillator (LO) is
sent into the resonator pumped optically (Fig.~\ref{fig1}). The RF
modulation frequency of the microwave carrier cannot exceed the
spectral width of both the optical and microwave resonances. The
pump light interacts with the microwaves creating optical harmonics
when the carrier frequencies of the signal and LO coincide with the
integers of the free spectral range (FSR) of the WGM resonator.
Each harmonic is spectrally separated from the optical carrier by a
value equal to a linear combination of the signal and local
oscillator frequency. The nonlinear process also changes the
transmission of the optical carrier. Information about the
microwave signal is retrieved by means of processing and detecting
the optical sidebands and$/$or the optical carrier leaving the
resonator.

\section{Experiment}

\begin{figure}[ht]
   \begin{center}
   \begin{tabular}{c}
   \includegraphics[height=6.5cm,angle=0]{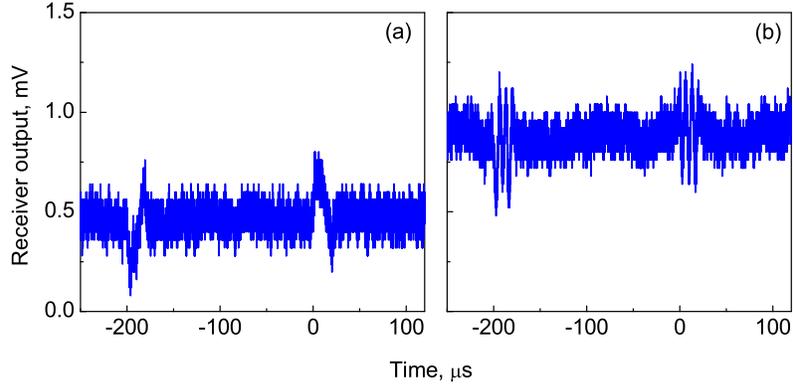}
   \end{tabular}
   \end{center}
   \caption[figs1]
{ \label{fig2}   The coherent receiver baseband response.
The optical coupling efficiency is approximately 60\%, and the optical insertion
loss away of the WGM resonance is less than $3$~dB. The input optical power of the receiver does
not exceed 50~$\mu$W. The signal pulses have
0.4~nW power and 20~$\mu$s duration. Their carrier frequency is approximately  35~GHz
(FSR of the WGM resonator).
Microwave local oscillator (LO) has 10~$\mu$W DC power and 50 kHz (a) [100 kHz (b)]
detuning from the signal carrier frequency. There is no time averaging of the response.
The fringes occur due to interference of the LO and the signal. The sensitivity of the
receiver corresponding to SNR$=$1 is better than 0.1~nW. The dynamic range is better than 50 dB.
The bandwidth of the receiver is
given by the optical linewidth of the WGMs and is approximately 5 MHz.
Measurement bandwidth (scope) is 20 MHz.
The gain of the receiver $G$ is equal to 0.01.
It is evaluated from the data assuming the detector resistance $\rho_{pd}=10$~kOhm and
amplitude of the retrieved signal $V=0.35$~mV. The power of the measured noise
approximately corresponds to the thermal noise power in the bandwidth of the measurement.}
   \end{figure}

In our experiment the lithium niobate WGM resonator with 35 GHz FSR
was excited using 1550 nm laser light at the resonant frequency of
one of the optical TE modes (the electric field is parallel to the
resonator symmetry axis which coincides with the c-axis of
LiNbO$_3$).  The laser frequency is kept at the center of the
optical resonance. The light output of the resonator received at a
low frequency photodetector reproduced the product of the low
frequency signal and the LO, both applied to the resonator. We have
applied 10 kHz to 100 kHz signals modulated on the 35 GHz microwave
carrier as well as a DC LO detuned 50 kHz to 500 kHz from the
signal carrier frequency to the resonator. A couple of examples of
the detected signal are shown in Fig.~(\ref{fig2}).

The operation principle of the coherent receiver is based on the RF
modulation of the resonant absorption of the pumping light when
both the microwave signal and LO are present. A critically coupled
resonator absorbs all the input resonant light, and the spectrum of
the modes of a critically coupled resonator has 100\% contrast.
When a microwave LO is applied to the resonator the excited WGM
resonance broadens and the light coupling efficiency decreases
\cite{ilchenko03josab,ilchenko08ptl}. First of all, the LO results
in generation of the optical sidebands that escape the resonator
and show overall increase of the optical power on the slow
photodiode. Microwave signal applied to the same resonator results
in time dependent change of the efficiency of the optical coupling.
The time dependent change of the optical power transmitted through
the resonator is proportional to the product of the amplitudes of
the signal and the LO. Hence, the photocurrent created by the light
exiting the resonator depends on the relative phase of the LO and
the signal as well as on the amplitudes of the signal and the LO.

\section{Theory}

Let us theoretically study the receiver properties. We consider the
nonlinear interaction of a WGM having resonant frequency $\omega_0$
and pumped with resonant light, as well as a microwave field mode
having frequency $\omega_{mw}$ and pumped with radiation having
frequency $\omega_M$. We study the generation of multiple Stokes
and anti-Stokes sidebands as the result of the nonlinearity in the
resonator medium, having $\omega_0-l\omega_M$ and $\omega_0 +
l\omega_M$ frequencies, where $l$ is an integer number. In what
follows we consider only the lowest sideband order $l=1$, neglect
by the dispersion of the resonator, and assume that the microwave
frequency $\omega_M$ is nearly equal to the free spectral range
$\omega_{FSR}$ of the resonator ($|\omega_M-\omega_{FSR}|\ll 1$) as
well as $\omega_M=\omega_{mw}$.

The power of the light escaping the resonator and generating the
base band signal can be presented as
\begin{eqnarray}
\frac{P_{out}|_{BB}}{P_{in}} &\simeq& \biggl | \sqrt{1-\xi^2} e^{i \phi_\xi}  + \xi \left (
1-\frac{2\gamma_c}{\Gamma_A} \frac{\Gamma_+ \Gamma_-}{\Gamma_+ \Gamma_-+2g^2 (t)} \right ) \biggr |^2 +
\xi^2 \left | \frac{2\gamma_c}{\Gamma_A} \right |^2 \frac{g^2(t) (|\Gamma_+|^2 + |\Gamma_-|^2)}{|\Gamma_+
\Gamma_-+2g^2 (t)|^2} \nonumber \\ \label{power}
\end{eqnarray}
where the first and the second terms in the right hand side of
Eq.~(\ref{power}) stand for the relative power at the carrier
frequency and at the sideband frequencies (here we take into
account the first sideband pair only),
$\Gamma_A=\gamma+\gamma_c+i(\omega_0-\omega)$ and
$\Gamma_{\pm}=\gamma+\gamma_c+i(\omega_0-\omega) \pm
i(\omega_{FSR}-\omega_M)$ are the tuning parameters for the carrier
and sideband modes, $\omega_0$ is the frequency of the pumped WGM,
$\omega$ is the frequency of the laser, $\gamma$ and $\gamma_c$ are
the intrinsic and coupling half widthes at half maximum of the WGM
resonances ($\gamma+\gamma_c$ is the loaded half width at  half
maximum), $\xi$ is the phase mismatch parameter for the coupling
prism and the resonator, $\phi_\xi$ is the phase of the part
optical signal that does not interact with the resonator, and
$g(t)$ is the modulation parameter given by
\begin{eqnarray} \label{s1}
g^2(t) =  \omega^2 r_{33}^2 n_e^4  \eta^2  \frac{\pi
 Q_M}{n_M^2 \omega_M {\cal V}_M} (P_{mw}+ P_{LO}+2\sqrt{P_{LO}P_{mw}} \cos
 (\omega_M-\omega_{LO} + \phi_M)).
\end{eqnarray}
Here, $Q_M$ is the quality factor of the critically loaded
microwave mode, $r_{33}$ is the electro-optical coefficient of the
resonator host material, $n_e$ and $n_M$ are the indexes of
refraction of the material, ${\cal V}_M$ is the volume of the
microwave field, $\eta = (1/V_e)\int |\Psi_e|^2\Psi_M dV <1$ is the
overlap integral of the optical and microwave fields, $|\Psi_e|^2$
and $|\Psi_M|^2$ are the spatial distributions of the power of the
optical and microwave fields respectively, $(1/V_e)\int |\Psi_e|^2
dV = 1$, $P_{mw}$ is the power of the microwave signal, $P_{LO}$ is
the power of the local oscillator, $\phi_{M}$ is the relative phase
of the local oscillator and the signal.

In the case of the resonant tuning of the laser as well as
quasi-critical ($\gamma_c=\gamma$ but $\xi$ is not necessary equal
to $1$) coupling the unsaturated signal is given by
\begin{equation} \label{linear}
\frac{P_{out}|_{BB}}{P_{in}} \approx \xi^2 \frac{2g^2(t)}{4\gamma^2+(\omega_{FSR}-\omega_M)^2}.
\end{equation}
Therefore, the microwave bandwidth of the receiver is equal to $4
\gamma$, which corresponds to the optical bandwidth.

It is convenient to introduce the microwave saturation power
$P_{sat}$ of the receiver showing when the response of the receiver
starts to decrease with the increase of the signal power
\begin{equation}
P_{sat} = \frac{n_M^2 \omega_M {\cal V}_M}{8\pi \eta^2 Q^2 Q_M r_{33}^2 n_e^4}.
\end{equation}
It corresponds to $\Gamma_+ \Gamma_-=2g^2 (t)|_{max}$ and allows to
estimate the maximum detectable microwave power which should not to
exceed significantly the saturation power in avoidance of the
decrease of the signal. The receiver is also characterized with the
gain $G$, which is given by the ratio of the power of the output
intermediate frequency (IF) signal and the input high frequency
signal. We assume that the DC part of the photocurrent generated by
the modulated light on the slow photodiode is filtered out. The AC
part of the photocurrent ($j={\cal R} P_{out}|_{BB}$) is
\begin{equation}
j_{AC}=2 {\cal R} \frac{ \xi^2 P_{in}}{P_{sat}} \frac{\sqrt{P_{LO}P_{mw}} \cos [(\omega_M-\omega_{LO})t +
\phi_M)]}{ 1+(\omega_{FSR}-\Omega_M)^2/(4\gamma^2)}  ,
\end{equation}
where ${\cal R}$ is the responsivity of the photodiode. The average
power of the AC signal is given by expression
\begin{equation}
P_{AC} = \xi^4  \frac{\rho_{pd} {\cal R}^2 P_{in}^2}{P_{sat}^2} P_{LO}P_{mw} \left [
1+\frac{(\omega_{FSR}-\Omega_M)^2}{4\gamma^2}  \right ]^{-2},
\end{equation}
where $\rho_{pd}$ is the resistivity of the photodiode. The gain of
the receiver is
\begin{equation}
G \equiv \frac{P_{AC}}{P_{mw}} =  \xi^4  \frac{\rho_{pd} {\cal R}^2 P_{in}^2}{P_{sat}}
\frac{P_{LO}}{P_{sat}}\left [ 1+\frac{(\omega_{FSR}-\Omega_M)^2}{4\gamma^2}  \right ]^{-2}.
\end{equation}

The output noise power of the receiver is given by the thermal
(input and the intrinsic noises) and relative intensity noises
\begin{equation} \label{small2}
P_{noise}= \left [ (1+G) k_BT + \rho_{pd} {\cal R}^2 P_{DC}^2 \frac{RIN}{\zeta} \right ]\Delta F,
\end{equation}
$k_B$ is the Boltzmann constant, $T$ is the temperature,  $\zeta <
1$ is the quantum efficiency of the photodiode, $\Delta F$ is the
video-bandwidth, and $P_{DC}$ is the DC part of the optical power
falling on the photodiode. The minimum of the RIN is given by
$RIN_{min} = 2 \hbar \omega/P_{DC}$. In a realistic receiver
$P_{DC} \simeq (1-\xi^2) P_{in}$ and the noise power is primarily
given by the RIN. In an ideal receiver $\xi \rightarrow 1$, $G \gg
1$, and the thermal noise determines the input RF noise power.

The noise floor power of the ideal receiver, determined as the
power that corresponds to unity signal to noise ratio ($S/N=1$) is
\begin{equation}
P_{mw\; noise \; floor}= \frac{P_{noise}}{G}
\end{equation}
The noise floor power and the saturation power determine the
dynamic range of the receiver
\begin{equation}
DR \approx \frac{P_{sat}}{P_{mw\; noise \; floor}}.
\end{equation}
In the ideal case $DR=P_{sat}/(k_BT\Delta F)$. The dynamic range
does not depend on either the optical power or the saturation power
of the receiver and is given by
\begin{equation}
DR \leq \frac{\xi^4}{(1-\xi^2)^2}  \frac{\zeta}{RIN \, \Delta F}
\end{equation}
when the noise of the receiver is determined by the RIN of the
laser, the signal frequency coincides with the FSR of the
resonator, and $P_{LO}=P_{sat}$. Low frequency RIN can reach
$-130$~dB$/$Hz for 3~mW output optical power for a standard DFB
laser. Assuming that $\Delta F=5$~MHz, $\xi^2=0.6$, $\zeta=0.8$, we
obtain $DR \leq 66$~dB.  On the other hand, the dynamic range of
the ideal receiver having $P_{sat}=10$~$\mu$W is $DR = 87$~dB. For
a good DFB laser RIN can be as low as $-160$~dB$/$Hz, so the noise
of such a laser is given by thermal noise, not by the RIN.
Increasing the saturation power leads to increase of the dynamic
range of the receiver. The receiver with $P_{sat}=1$~mW can have
107~dB dynamic range.

In detection of deterministic narrowband signals where optimal
filtration or postfiltration can be applied the effective noise
bandwidth of the receiver can be reduced several orders of
magnitude. Such a reduction will lead to a significant increase of
the dynamic range as well as the sensitivity of the receiver.

Let us theoretically estimate the parameters of the experimental
realization of the coherent receiver reported above. We assume that
$Q=4 \times 10^7$, so that FWHM of the optical resonance is
$5$~MHz, $\Delta F = 5$~MHz, and $\gamma=\gamma_c=2\pi\times 1.3
\times 10^6$~s$^{-1}$. We further assume $Q_M=60$, $\omega_M = 2\pi
\times 35$~GHz, $P_{in}=50$~$\mu$W,
$r_{33}=30$~pm$/$V=$7\times10^{-8}$~esu, $n_e=2.1$, $n_M=5.4$,
${\cal V}_M = 3 \times 10^{-6}$~cm$^3$, $\rho_{pd}=10$~kOhm, ${\cal
R}=0.8$~A$/$W, $k_B=1.38\times 10^{-23}$~J$/$K. Using these values
we obtain $G=0.6$, $P_{mw\ sat}=10$~$\mu$W, $P_{mw\; noise \;
floor}= 3$~pW, and $DR=65$~dB. In the experiment reported in
Fig.~(\ref{fig2}) we have lower gain because of the optical and
microwave losses throughout the system.

\section{Discussion}

To improve the sensitivity as well as dynamic range of the receiver
we need to increase the optical power retrieved out of the
resonator as well as to reduce the laser noise. Locking of the
laser to the resonator is very promising because it
results in a decrease of both low frequency RIN and phase noise of
the laser  \cite{li89jqe}. Moreover, locking allows keeping the
laser frequency exactly at the center of the WGM, so the
transformation of the phase to amplitude laser noise is
significantly suppressed. The locking can be realized, e.g., using
Rayleigh scattering \cite{vassiliev98oc}.

In our experiment we were unable to increase the optical power
beyond 50~$\mu$W because of the optical damage of lithium niobate
as well as mode interactions in the resonator
\cite{savchenkov07oc}. We have found that the optical damage is
much lower in lithium tantalate resonators. The power can also be
increased if one reduces the quality factor of the resonator. Such
a reduction leads to the larger bandwidth of the receiver on the
one hand, and lower saturation power on the other. Usage of
single-mode resonators \cite{savchenkov06ol} will also lead to a
significant reduction of the mode interaction because of the
elimination of the unwanted modes.

Let us now compare the coherent and quadratic
\cite{hosseinzadeh05sse,hosseinzadeh06tmtt,ilchenko08ptl}
receivers. According to the results of the experiments reported in
\cite{ilchenko08ptl} and this paper, we find that the dynamic range
of the coherent receiver is at least 10~dB larger than the dynamic
range of the quadratic receiver. According to our theoretical
calculations the dynamic range of the coherent receiver is 20~dB
higher for the same experimental conditions and can potentially be
40~dB higher for 1~mW optical pump power. The sensitivity of the
coherent receiver is also much higher than the sensitivity of the
quadratic receiver. Moreover, the coherent receiver allows
measuring the phase and frequency of the microwave signal, along
with its amplitude.

Resonant quadratic photonic receivers
\cite{hosseinzadeh05sse,hosseinzadeh06tmtt,ilchenko08ptl} are
fundamentally nonlinear. A major problem with them is the
generation of second-order products of signals from two nearby RF
channels. One component of these second-order products is located
at base band, and thus interferes with the desired signal,
degrading performance. The coherent receiver we discuss here is
linear due to presence of the strong LO. It can be used for
reception of several signals coming from a single channel.

The performance of the linear receivers is generally given by the
sensitivity, spurious-free dynamic range, and third-order intercept
point. Spurious-free dynamic range and third-order intercept point
depend on the residual cubic nonlinearity of the receiver. The
method for characterizing a linear receiver is based on injection
of equal levels of two closely spaced carriers into the device and
the study of generated spurious sidebands at product frequencies.
Our coherent photonic receiver does not possess a significant cubic
nonlinearity, and the nonlinearity of the photodiode as well as the
RF circuitry can be quite small.  Thus, determinations of a third
order intercept point as a measure of signal distortion is not
applicable for the device. Stated more succinctly, our photonic
receiver does not produce any distortion resulting from third order
nonlinearity by itself.

On the other hand, the receiver does produce distortion as a result
of the second order nonlinearity.  But as is well known, all
distortions resulting from the second order nonlinearity occur at
frequencies that are widely separated from the input frequency.
Consequently there are no distortion of the closely separated
signals. For example, if the receiver has $35$~GHz LO and two
signals with carrier frequencies $35.001$~GHz and $35.0011$~GHz fed
into the RF resonator, the receiver shifts frequencies of those
signals to $1$~MHz and $1.1$~MHz in the base band. The second order
nonlinearity results in the spurious sideband at $100$~kHz
frequency. This low frequency sideband can be easily filtered out. As the rule, if the LO frequency is separated with the receiver band by several reception bandwidths, the second order nonlinearity does not degrade the receiver performance. If the LO frequency located within the reception band, the second order nonlinearity limits the spurious free dynamic range of the
receiver to approximately 50 dB.

\section{Conclusion}

We have demonstrated experimentally and studied theoretically an
all-resonant coherent photonic microwave receiver based on a
lithium niobate whispering gallery mode resonator. The receiver has
high sensitivity and large dynamic range, and its performance does
not degrade for higher microwave frequencies. Hence, the photonic
receiver is an attractive alternative to the conventional
electronic receivers and will likely enable new applications.

\end{document}